\documentclass[sigconf]{acmart}
\makeatletter
\def\@ACM@checkaffil{% Only warnings
    \if@ACM@instpresent\else
    \ClassWarningNoLine{\@classname}{No institution present for an affiliation}%
    \fi
    \if@ACM@citypresent\else
    \ClassWarningNoLine{\@classname}{No city present for an affiliation}%
    \fi
    \if@ACM@countrypresent\else
        \ClassWarningNoLine{\@classname}{No country present for an affiliation}%
    \fi
}
\makeatother
\usepackage{subcaption}
\usepackage{algorithm}
\usepackage{algorithmic}

\AtBeginDocument{%
  \providecommand\BibTeX{{%
    \normalfont B\kern-0.5em{\scshape i\kern-0.25em b}\kern-0.8em\TeX}}}

\setcopyright{acmcopyright}
\copyrightyear{2023}
\acmYear{2023}
\acmDOI{XXXXXXX.XXXXXXX}

\acmBooktitle{EDBT/ICDT 2024 Joint Conference,
 25th - 28th March, 2024, Paestum, Italy}
\acmPrice{15.00}
\acmISBN{978-1-4503-XXXX-X/18/06}

\setcopyright{none}
\renewcommand\footnotetextcopyrightpermission[1]{} % removes footnote with conference information in first column
\begin{document}

%%
%% The "title" command has an optional parameter,
%% allowing the author to define a "short title" to be used in page headers.
\title[BitE : Accelerating Learned Query Optimization in a Mixed-Workload Environment]{BitE : Accelerating Learned Query Optimization in a Mixed-Workload Environment}

%%
%% The "author" command and its associated commands are used to define
%% the authors and their affiliations.
%% Of note is the shared affiliation of the first two authors, and the
%% "authornote" and "authornotemark" commands
%% used to denote shared contribution to the research.
\author{Yuri Kim}
\authornote{indicates equal contribution. This work was done when Yuri Kim, Yewon Choi and Yujung Gil were interns in SAP Labs Korea}
\email{yuri.kim@sap.com}
\orcid{0009-0003-3338-3749}
\affiliation{%
  \institution{Korea University}
  }
\affiliation{%
  \institution{SAP Labs Korea}
  \city{Seoul}
  \country{Korea}
}

\author{Yewon Choi}
\email{yewon.choi@sap.com}
\orcid{0009-0003-3738-3738}
\authornotemark[1]
\affiliation{%
  \institution{Ewha Womans University}
  }
  \affiliation{%
  \institution{SAP Labs Korea}
  \city{Seoul}
  \country{Korea}
}

\author{Yujung Gil}
\authornotemark[1]
\email{yujung.gil@sap.com}
\affiliation{%
  \institution{Dongguk University}
}
\affiliation{%
  \institution{SAP Labs Korea}
  \city{Seoul}
  \country{Korea}
}

\author{Sanghee Lee}
\email{sanghee.lee@sap.com}
\affiliation{%
  \institution{SAP Labs Korea}
  \city{Seoul}
  \country{Korea}
}

\author{Heesik Shin}
\email{heesik.shin@sap.com}
\affiliation{%
  \institution{SAP Labs Korea}
  \city{Seoul}
  \country{Korea}
}

\author{Jaehyok Chong}
\authornote{Corresponding author.}
\email{ja.chong@sap.com}
\affiliation{%
  \institution{SAP Labs Korea}
  \city{Seoul}
  \country{Korea}
}

%%
%% By default, the full list of authors will be used in the page
%% headers. Often, this list is too long, and will overlap
%% other information printed in the page headers. This command allows
%% the author to define a more concise list
%% of authors' names for this purpose.
% \renewcommand{\shortauthors}{Trovato and Tobin, et al.}
\renewcommand{\shortauthors}{Kim and Choi, et al.}

%%
%% The abstract is a short summary of the work to be presented in the
%% article.
\begin{abstract}
Although the many efforts to apply deep reinforcement learning to query optimization in recent years, there remains room for improvement as query optimizers are complex entities that require hand-designed tuning of workloads and datasets. Recent research present learned query optimizations results mostly in bulks of single workloads which focus on picking up the unique traits of the specific workload. This proves to be problematic in scenarios where the different characteristics of multiple workloads and datasets are to be mixed and learned together. Henceforth, in this paper, we propose BitE, a novel ensemble learning model using database statistics and metadata to tune a learned query optimizer for enhancing performance. On the way, we introduce multiple revisions to solve several challenges: we extend the search space for the optimal Abstract SQL Plan(represented as a JSON object called ASP) by expanding hintsets, we steer the model away from the default plans that may be biased by configuring the experience with all unique plans of queries, and we deviate from the traditional loss functions and choose an alternative method to cope with underestimation and overestimation of reward. Our model achieves 19.6\% more improved queries and 15.8\% less regressed queries compared to the existing traditional methods whilst using a comparable level of resources.   
\end{abstract}

%%
%% The code below is generated by the tool at http://dl.acm.org/ccs.cfm.
%% Please copy and paste the code instead of the example below.
%%

%%
%% Keywords. The author(s) should pick words that accurately describe
%% the work being presented. Separate the keywords with commas.

% \received{1 June 2023}
% \received[revised]{12 March 2009}
% \received[accepted]{5 June 2009}

%%
%% This command processes the author and affiliation and title
%% information and builds the first part of the formatted document.
\maketitle
\pagestyle{plain}

\section{Introduction} \label{intro}
Recently, there have been many attempts to apply machine learning techniques to database systems\cite{b1, b2}, especially to query optimization. Diverse approaches exist depending on various factors, for instance, the learning methods (e.g. supervised, reinforcement), algorithms of the model (e.g. CNN, decision tree), and the target of the value network (e.g. cardinality, cost). However, to the best of our knowledge, we noticed that recent research work is still limited in that it greatly focuses on a learning model \textit{per workload}\cite{b3, b4, b5, b6, b7, b8, b9} and poses several discussion points concerning the efficiency of query optimization in mixed-workload environments. We mainly tested our hypothesis and expanded on Bao\cite{b4} as our \textit{baseline model} in connection to the \textit{expert optimizer}(HANA), a high-speed in-memory database.

\textbf{(1) Expandability} If the value networks are dependent on workloads, newly introduced workloads will not be able to be integrated right away as their performance will not match any of the original models and therefore a new model will have to be set up for it. This strays from our objective to construct a more generalized version of a query optimizer. 

\textbf{(2) Expensive to build and manage} In order to construct individual models for each workload, extra resources are exhausted to train and pertain the model in areas such as memory and GPU usage(See Section \ref{ensemble-results-section}). In addition, as the pretrained models are retrained regularly to optimize performance when user input queries are accumulated, a repetitive process will occur, retraining take place for multiple models. 

\textbf{(3) Limited Performance}
In most of the aforementioned applications of ML for query optimization, metrics for evaluation of performance are provided individually on separate workloads without being mixed together or otherwise were focused only on the join problems. While many of the ML models show impressive accomplishment in single-workload driven query optimization, it begs the question, is the performance also sustainable in environments that take in multiple workloads at once?

\begin{figure}[htbp]
\centerline{\includegraphics[width=9cm]{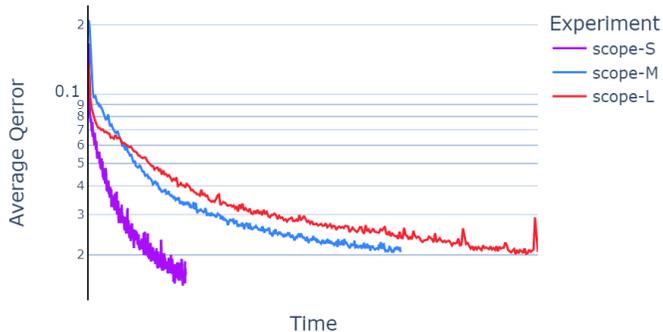}}
\caption{Q-error for never-before-seen queries. Q-error shows the loss of the model. Scope-S, M, L denotes the experiments that have differing number of workloads used as training data, in ascending order. Scope-S which is based on a single workload shows significantly better performance compared to Scope-L which uses 9 workloads at once.}
\label{qerror-figure}
\end{figure}

To this end, we went over a series of experiments and learned that performance really does show correlation with the diversity in workloads(Figure \ref{qerror-figure}, further elaborated in Section \ref{hypothesis}). We propose a plausible solution of \textbf{BitE}(Bias-tuning Ensemble Model) that utilizes the complexity of database statistics to classify models. BitE labels the workloads into two classifications - Light and Heavy - and shape them into a two-way ensemble model based on the complexity characteristics of the workloads. While the complexity and weight of the execution of a query depend on many elements, we mainly focus on three aspects of workloads to determine the classification of each workload. 

\begin{itemize}

\item \textbf{Scale factor} : Workloads such as TPCH and TPCDS have scale factors which stand for the predetermined size of the tables data. In the case of TPC-DS benchmark, fact tables which are the tables with the biggest number of records determine the scale factor. 

\item \textbf{Skewness} : Skew in filter, aggregation and join operations make query optimization more challenging. In this category, the JCCH workload introduces correlations and skewness from the original TPCH queries so that filter predicates alters the frequency and join-fan-out-distributions of the plan operators. 

\item \textbf{Magnitude of tables} : Apart from scale factors, the general magnitude of the workload also affects query execution. Even for JOB, JOB-m, JOB-e, JOB-light, JOB-light-ranges benchmarks that mainly root from the same IMDB workload, the number of tables, columns and maximum records of a table actually used and referred to differ.
\end{itemize}

This way, the proposed method would address the three limitations that the individual workload models have. 

\textbf{(1) Expandable for new workloads} Even when unseen workloads are put through our model, only a simple stage of labeling the workload to either Light or Heavy would be required in a similar manner using statistical information of the workload. Then, instead of training an entire model which could take hours, the classified workload or set of queries could be used immediately for query optimization and hintset prediction. 

\textbf{(2) Untroublesome to build and manage} Rather than having infinite sets of models accounting for each workload, we would only need to maintain two networks to be later on put in our ensemble model. No matter how many workloads we work with, the number of classifications remains the same. 

\begin{figure}[htbp]
\centerline{\includegraphics[width=9cm]{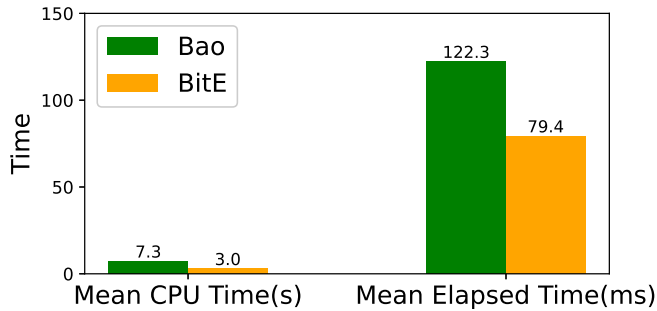}}
\caption{Comparison between baseline model(Bao) and BitE of mean execution time for a subset of target queries that do not fall in the scope of timeouts. CPU and elapsed time are each in the unit of seconds and milliseconds.}
\label{cpu-total-figure}
\end{figure}

\textbf{(3) Better performance with multiple workloads} We compare the mean taken CPU and elapsed time to execute approximately 1500 target queries, each using the plans generated by our baseline model and BitE, which takes into account the complexity of different workloads. Based on Figure \ref{cpu-total-figure}, BitE shows 59\% and 35\% improvement each in total CPU and elapsed execution time, \textit{far outperforming} the baseline model in its efficiency in mixed-workload environments. \\

We also introduce additional key contributions to accelerate query optimization of BitE in the upcoming sections: 
 
\textbf{Expanding hintsets.} Since the expert optimizer targets and supports hundreds of logical enumeration hints\cite{b10}, with full utilization, the search space of the model becomes extremely large with  $2^n$ cases for $n$ hints(e.g. For 100 hints, $2^{100}$ hintsets are created as each hints can be turned on or off). Therefore it is important to select and compress effective hintsets to reduce our search space. We group hints that are frequently used together and at the same time consider logical sense of the sets for efficiency(Section \ref{expanding-hintsets}). In a sense, we find the optimum numbers of hintsets to fully take advantage of and expand the search space for optimal query plans but at the same time, limit hintsets so that the overhead of executing hintsets for each query doesn’t grow too big as a burden on our system. To control this overhead in growing number of hintsets, we later introduce parallel compilation (Section \ref{soft-landing-section}) and integrate it into our learning method.

\textbf{Removing bias in query optimization.} In the original experiment methods of the baseline model, it collects initial data with no-hint plans, that is the default plans generated from the expert optimizer, which can invoke bias of the model to prefer no-hint. Even though no-hint plans will be safe enough since it relies on the expert optimizer, our goal is finding ‘new’ plans that perform better than the traditional optimizer. So, we construct experiments to remove the bias and find more of the new unbiased, but more efficient plans (Section \ref{Entire Hintsets}). To eliminate the bias of learning according to the pattern of the plan preferred by the expert optimizer, we accumulate initial training data by generating all plans from every hintset instead of only generating ‘no-hint’ plans. We also replace Q-error as our new training loss to give all queries equal weight in forming BitE, eliminating the bias towards heavy queries(Section \ref{qerror-bias-section}).

\section{Implementing Ensemble Models for Multiple Workloads} \label{bite_intro}
Even though Learned models are not intended to learn logical enumeration patterns workload-dependently, through a series of experiments, we noticed that the more tables are brought in the dataset, the worse the predictability becomes. Hence we set up a series of experiments to prove if the diversity in workloads really do show correlation with query performance in Section \ref{hypothesis}. We then go into how to identify the complexity of the workloads using the statistical information in Section \ref{ensemble-interpretation}. Finally, Section \ref{system architecture} illustrates the specifics of implementing BitE which we proposed in Section \ref{intro}, a plausible solution of measuring the complexity of the individual workloads to later group them together to form a ensemble model.

\subsection{Regression in Mixed-Workloads Environment} \label{hypothesis}

\begin{table}[]
\begin{tabular}{|l|l|l|l|}
\hline

\textbf{workload} & \multicolumn{1}{c|}{\textbf{\# Queries}} & \multicolumn{1}{c|}{\textbf{\begin{tabular}[c]{@{}c@{}}Cumulative \\ Total\end{tabular}}} & \multicolumn{1}{c|}{\textbf{Experiment set}} \\ \hline
JOB               & 100                                             & 100                                            & \begin{tabular}[c]{@{}l@{}}scope-S, scope-M,\\ scope-L\end{tabular}                    \\ \hline
TPCDS\_SF1        & 73                                              & 173                                            & scope-M, scope-L                             \\ \hline
TPCDS\_SF10       & 28                                              & 201                                            & scope-M, scope-L                             \\ \hline
STACK             & 99                                              & 300                                            & scope-M, scope-L                             \\ \hline
TPCH\_SF1         & 22                                              & 322                                            & scope-L                                      \\ \hline
TPCH \_SF10       & 22                                              & 344                                            & scope-L                                      \\ \hline
TPCH\_SF100       & 22                                              & 366                                            & scope-L                                      \\ \hline
JCCH              & 22                                              & 388                                            & scope-L                                      \\ \hline
TPCDS\_SF10       & 50                                              & 438                                            & scope-L                                      \\ \hline
Corporate         & 62                                              & 500                                            & scope-L                                      \\ \hline
\end{tabular}
\newline
\caption{Number of queries each used to train the scope experiment models. The experiment set name signifies the target coverage of various workload- L for large, M for medium and S for small- and which workloads were included in it. 50 more TPCDS-10 queries were extracted to extend the magnitude of scope-L. Corporate is an internal dataset and added to diversify the workloads for this experiment with 78 tables and 254M records for the largest table.}
\label{3experiments-table}
\end{table}

\begin{table}[]
 \centering
\begin{tabular}{|l|l|l|l|}
\hline
\textbf{Experiment set} & \multicolumn{1}{c|}{\textbf{\# Improved}} & \multicolumn{1}{c|}{\textbf{\# Regressed}} & \multicolumn{1}{c|}{\textbf{\# Workloads}} \\ \hline
Scope-S                 & 32                                        & 7                                          & 1                                          \\ \hline
Scope-M                 & 24                                        & 9                                          & 4                                          \\ \hline
Scope-L                 & 23                                        & 10                                         & 9                                          \\ \hline
\end{tabular}
\newline
\caption{Results of experiment predictions. Evaluation performed upon 100 queries from the IMDB dataset}
\label{sml-results-table}
\end{table}

We set up three different workloads each for experiments to see if the variety of workloads affect performance, each experiment set being incremental in the number of workloads. Each experiment scope-S, scope-M, scope-L extracted 100, 300, 500 queries each from 1, 4 and 9 workloads, described in Table \ref{3experiments-table}.

For each experiment set, we measured whether the predictions improved for 100 target queries. As shown in Table \ref{sml-results-table}, the improvements were induced in comparison with the execution of the queries with no hints, that is the default optimal plan that the expert optimizer figures. With one workload used solely as the training set, Scope-S had the most improved queries and at the same time, the least regressed queries. As the workloads covered by the different experiment sets grew, there were less improved queries and more regressed queries. This served as evidence that clearly, the efficacy decreases for multiple workloads as the models are tuned in multiple directions. 

For experiments with bigger loads of datasets of workloads as shown in Figure \ref{qerror-figure}, which should enhance the prediction accuracy by providing the model with more patterns to learn, the Q-error(loss for model, Section \ref{qerror-bias-section}) seemed to be bigger and slower to converge. Scope-L which had the most workloads showed a greater variation even during training compared to the other sets, indicating that with diverse workloads, it suffered from the danger of serious regressions if the training is not stopped at the appropriate stage. Therefore, we cope with these spikes in mixed-workload environments by proposing BitE.

\subsection{Accounting for Workload Complexity} \label{ensemble-interpretation}
As proposed in Section \ref{intro}, we try to solve the problem by dividing the model into two versions based on the statistical information of the workload - Light and Heavy.
This way, when a new workload query comes in, it can be classified as either Light or Heavy and passed through the corresponding model based on the statistical information, which doesn’t require any additional model setup.

As mentioned in Section \ref{intro}, we turn to scale factors, skewness and the magnitude of the workloads to determine the complexity weight of the workload. For this matter, Table \ref{workloads-table} includes a detailed table statistics for every workload. 

\textbf{Why use table statistics and metadata to classify workloads and how are they related to query optimization?}

Although there exist attempts to provide a more generalized model for query optimization\cite{b21}, we observed that most research work focus on a single-workload environment as some traits of the plan are unique to workloads and can’t be disconnected from the workload itself. In addition, another factor to consider is that the magnitude or complexity of workloads are a determining factor of execution time, as they establish the cost and size of queries. Hence if we distinguish the characteristics of the workload and group it together to create a ensemble model, this would preserve the workload information and help create an efficient, high-performing model. In addition, the action of manually classifying the complexity of each workload could provide to the fact that cost and size estimations are often miscalculated as size depends on a set of predefined rules with an assumption of data distribution and would contribute to reducing the impact of the misguided estimations by providing an general sense of how heavy the execution will be. 

However, as there are several standards to tend to, there is no “one clear” rule to follow when distinguishing the type of allotment. Henceforth, we move on to find the best combination of workloads to achieve this in a rather heuristic manner by trying out the options carefully designed and selected. We also wanted to compare using metadata of the tables and the size of the latency we experienced for the division while preparing for experiments and the weight to see if the metadata is an accurate and optimal way of constructing the model. Hence for Ensemble 1 and Ensemble 2, the separation was based more on the CPU execution time while Ensemble 3 and Ensemble 4 was purely based on the three factors mentioned above – scale factors, skewness and magnitude of tables.

\begin{table}[htbp]
\footnotesize
\begin{tabular}{lllllll}
\cline{1-6}
\multicolumn{1}{|l|}{\textbf{Experiment}} & \multicolumn{1}{c|}{\textbf{JOB}}   & \multicolumn{1}{c|}{\textbf{JOB\_m}} & \multicolumn{1}{c|}{\textbf{JOB\_light}} & \multicolumn{1}{c|}{\textbf{JOB\_e}} & \multicolumn{1}{c|}{\textbf{JCCH}}   &                               \\ \cline{1-6}
Ensemble 1                                & Heavy                               & Heavy                                & Light                                    & Heavy                                & Heavy                                &                               \\
Ensemble 2                                & Light                               & Light                                & Light                                    & Light                                & Heavy                                &                               \\
Ensemble 3                                & Light                               & Light                                & Light                                    & Light                                & Heavy                                &                               \\
Ensemble 4                                & Heavy                               & Heavy                                & Heavy                                    & Heavy                                & Heavy                                &                               \\
                                          &                                     &                                      &                                          &                                      &                                      &                               \\ \cline{1-6}
\multicolumn{1}{|l|}{\textbf{Experiment}} & \multicolumn{1}{l|}{\textbf{STACK}} & \multicolumn{1}{l|}{\textbf{TPCDS1}} & \multicolumn{1}{l|}{\textbf{TPCDS10}}    & \multicolumn{1}{l|}{\textbf{TPCH1}}  & \multicolumn{1}{l|}{\textbf{TPCH10}} &                               \\ \cline{1-6}
Ensemble 1                                & Heavy                               & Light                                & Heavy                                    & Light                                & Heavy                                &                               \\
Ensemble 2                                & Heavy                               & Heavy                                & Heavy                                    & Heavy                                & Heavy                                &                               \\
Ensemble 3                                & Heavy                               & Light                                & Light                                    & Light                                & Heavy                                &                               \\
Ensemble 4                                & Heavy                               & Light                                & Light                                    & Light                                & Heavy                                &                               \\
                                          &                                     &                                      &                                          &                                      &                                      &                               \\ \cline{1-4}
\multicolumn{1}{|l|}{\textbf{Experiment}} & \multicolumn{2}{c|}{\textbf{JOB\_light\_ranges}}                           & \multicolumn{1}{l|}{\textbf{TPCH100}}    & \textbf{}                            & \multicolumn{1}{c}{\textbf{}}        & \multicolumn{1}{c}{\textbf{}} \\ \cline{1-4}
Ensemble 1                                & Light                               &                                      & Heavy                                    &                                      &                                      &                               \\
Ensemble 2                                & Light                               &                                      & Heavy                                    &                                      &                                      &                               \\
Ensemble 3                                & Light                               &                                      & Heavy                                    &                                      &                                      &                               \\
Ensemble 4                                & Heavy                               &                                      & Heavy                                    &                                      &                                      &                              
\end{tabular}
\caption{Splitting workloads to Light and Heavy. }
\label{split-table}
\end{table}

For example, for experiment Ensemble 3, the division of the workloads into Light and Heavy are based on the scale factor, skewness of the data distribution and the magnitude of tables, that is the number of tuples of the biggest table. All four of the Heavy workloads have 60M records in the biggest table within the schema while JCCH adds on with the fact that it proves to be Heavy as its skewness acts as a hinderance to query optimization. The size and complexity of the workload would lead on to having a much more intricate and deeper neural network than the other, comparably simple “light” workloads and hence would result in better performance when it comes to query optimization. 

\begin{algorithm} \label{alg1}
\caption{Train Ensemble model }\label{alg1}
\hspace*{\algorithmicindent} \textbf{Input} Light workload queries, Heavy workload queries \\
\hspace*{\algorithmicindent} \textbf{Output} Learned model 
\begin{algorithmic}[1]
\STATE Construct(model\_classifier) \hfill\COMMENT{Add linear classifier at the forefront to classify the workload complexity with statistics}
\STATE
\FOR{\texttt{EnsembleModel\_class $\in$ \(Heavy, Light\) }}
    \FOR{\texttt{query $Q$ $\in$ model\_class workload }}
        \STATE \texttt{asps ← parallel\_compilation(Q)} \hfill\COMMENT{Extract all 225 abstract sql plans for all hintsets}
        \STATE \texttt{asps ← unique(asps)} \hfill\COMMENT{Remove duplicate plans}
        \FOR{\texttt{asp in asps}}
            \STATE \texttt{reward ← execute(asp)} \hfill\COMMENT{Execute plan and get execution time} 
        \ENDFOR
    \STATE \texttt{Experience ← save(reward, asp)}\hfill\COMMENT{Insert reward into Experience along with plan}
    \ENDFOR
\ENDFOR
\STATE \texttt{{Train(EnsembleModel\_class)}}
\STATE \texttt{return EnsembleModel\_class}
\end{algorithmic}
\end{algorithm}

 \begin{algorithm} \label{alg2}
\caption{Predict hintset for query with Ensemble model}\label{alg2}
\hspace*{\algorithmicindent} \textbf{Input} Query $Q$\\
\hspace*{\algorithmicindent} \textbf{Output} Hintset that returns optimal performance for the given query execution 
\begin{algorithmic}[1]
\STATE \texttt{Ensemble ←Classify($Workload_Q$)} \hfill\COMMENT{Classify the workload the query is based on to either Light or Heavy and load the according Ensemble model}
\STATE
\FOR{hintset $\in$ range(0, 225)}
    \STATE \texttt{$rewards_{hintset}$ ← Ensemble($Q_{hintset}$)} \hfill\COMMENT{Predict reward for every hintset with the Ensemble model}
\ENDFOR
\STATE
\STATE \texttt{Optimal Hintset ← argmin(rewards)} \\
\STATE return \texttt{Optimal Hintset}
\end{algorithmic}
\end{algorithm}

In Section \ref{ensemble-results-section}, we present and evaluate our proposal and our experimental runs, confirming the efficiency of the method.

\subsection{System Architecture} \label{system architecture}

\begin{figure}[htbp]
\centerline{\includegraphics[width=0.5\textwidth]{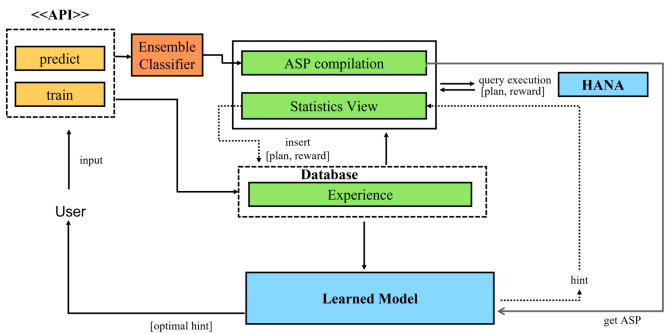}}
\caption{System architecture}
\label{fig1}
\end{figure}

In this phase we go over the architecture overview by discussing the process of how BitE works side to side with the expert optimizer, and showing how we extract the necessary components to the final evaluation of each model. The detailed architecture is shown in Figure \ref{fig1}.

\textbf{Training} The main reason for choosing Bao as our baseline model among other learned query optimizers in this paper is that it takes advantage of fully utilizing the knowledge of underlying expert optimizer. It is robust to changes in schema, data, or workload so there is no need to train from scratch when changes in the workload itself occur. Also, no modification to the expert optimizer is required while integrating the baseline model into it, while all features and knowledge supported by the underlying optimizer can be utilized. Our model is initially trained by queries generated with all predefined hintsets and retrained when new user queries come into the system(Algorithm \ref{alg1}). 

\textbf{Prediction} Input queries are first classified by our pre-built classifier to decide which model between Light and Heavy that the query will be passed into(Algorithm \ref{alg2}). The expert optimizer then compiles the query with all target hintsets and returns the abstract sql plan for each hintset through a built-in procedure. Returned ASPs represent estimated size, cost, and metadata information for the relationship of each operator. These ASPs of hintsets are put into learned model as an input. Once it passes the network, the model - constructed with several layers of tree CNNs, normalization as well as a pooling, fully connected and an activation layer - learns the affiliated patterns and costs existing in the plan. It returns a predicted value for the reward(i.e. the target of our learned model - CPU execution time) of each target hintset and of all the predicted rewards, the model picks the hintset with the optimal reward which should be the lowest execution time. We refer to this as the optimal predicted hintset. This selected hint is then appended to the query using the existing optimizer, and the resulting reward is stored in the experience, providing feedback to the model. Unless mentioned otherwise, CPU time is defined as our reward which is stored in an internal statistics view.  

\textbf{Evaluation metrics and sources} Once we collect enough experience to train our model, we start the training and evaluation. Evaluation is carried out on never-before-seen query $k_{n+1}$. Essentially, when a new query comes in, it is passed through the network, trained only with the initial queries, to predict the reward. It is not until this process is over that we actually calculate the loss to give feedback to the model through backpropagation, which constructs a model with +1 queries. This way, it is much more efficient and provides a timelier manner to evaluate our model, instead of having a separate set of validation queries to be executed every epoch or model to get the real CPU time which should be costly as well. 

\section{Accelerating Query Optimization}
In this section, we aim to describe experiments conducted to further tune and accelerate BitE for better performance. To achieve this, we identify and mitigate the limitations associated with the narrow hintset search space(Section \ref{expanding-hintsets}) and the two types of bias that exist in our baseline learned model(Section \ref{Entire Hintsets} and \ref{qerror-bias-section}).  

\subsection{Expanding hintsets} \label{expanding-hintsets}

Expanding hintsets from the minimum set of join problems, which many commonly approach query optimization with, is a crucial step to query optimization as it provides a sufficient search space. However, this must be dealt with caution as if it is overdone, the number of cases could grow exponentially and rather be detrimental to the learning steps.

Out of the hundreds of hints of the expert optimizer, we mark 15 significant hints (Table \ref{hints-table}) and combine them to come up with meaningful hintsets that will lead to potential optimal plans. 15 logical enumeration hintsets were chosen as distinct hints that will affect query optimization, and with each hint, there exists an option to turn it either on or off. However, if we just naively apply this to generate hintsets, it would result in $2^{15}$(more than 30,000) bandits which would not be feasible to compile all in real time. Therefore, we pick out and exclude combinations of hints that logically do not make sense. For example,  

\textbf{(1) hash\_join, index\_join}

It is logically impossible for the expert optimizer to choose both join predicates. They are mutually exclusive; therefore, one should be off while the other one is turned on. However, we do not eliminate the hintset of both being turned off, as another join predicate can be chosen instead of the pre-selected two. 

\textbf{(2) [range\_join, hashed\_range\_join], \newline \hspace*{0.75cm} [hash\_join, index\_join]}

The former two join predicate groups are applied only in queries with inequality (greater than, less than, or between) predicates, comparing the values of columns. When there are no inequality filters in queries, the latter two will mostly be exercised. Therefore, we place the hints into two independent groups and henceforth reduce the combinations of hints to half of the possible combinations. \\

$set(range\_join,hashed\_range\_join,hash\_join,index\_join)=\\
set(range\_join,hashed\_range\_join) \cup set(hash\_join,index\_join)$ \\

\textbf{(3) join\_thru\_aggr, aggr\_thru\_join}

The hints differ only in the order of logical enumeration plans. Hence when they come together, they bear no meaning as they are contrastive to each other.

\begin{table}[]
\centering
\begin{tabular}{|l|l|l|}
\hline
Category               & \multicolumn{1}{c|}{\# Hints} & \multicolumn{1}{c|}{Examples of hints}                                            \\ \hline
Joins(Enumeration) & 4                             & \begin{tabular}[c]{@{}l@{}}HASH\_JOIN \\ RANGE\_JOIN\end{tabular}       \\ \hline
Logical Enumeration    & 11                            & \begin{tabular}[c]{@{}l@{}}AGGR\_BEFORE\_UNION \\ FILTER\_THRU\_JOIN\end{tabular} \\ \hline
\end{tabular}
\newline
\caption{Hints used to build our learned model. }
\label{hints-table}
\end{table}

We end up with 225 hintsets, reducing the number of bandits to under 1\% of the original targeted combinations of hints. The number of hintsets are critical to query optimization as compilation is required for every single hintset - which would result in great latency if it were to grow too large. We found these to be the optimal number of hintsets: If too many, the latency would become too large and if too little, it would not provide a sufficient search space to find the optimal hintset. 

\subsection{Removing Bias with Entire Hintsets} \label{Entire Hintsets}

Originally, the baseline model undergoes an initial chunk where it goes through a process of model initialization. This process involves executing queries with ‘no-hint’ and in the n-chunk approach afterwards, the process involves iteratively sampling a subset of queries for prediction and retraining. 

However, there exists a drawback in this approach. If the model initialization is performed by accumulating experience using the default expert optimizer in the first chunk, it can lead to a bias, leaning towards the original plans generated by the expert optimizer. When bias occurs, there is a higher likelihood of selecting hints according to the expert optimizer's preferred methods, which may neglect the possibility of better plans in alternative spaces. To address this issue, we have devised the following method. 

In the first chunk, we introduce the ‘Unbiased’ method, a modified model initialization process where each sampled query's execution was accompanied by appending all 225 predefined hints. This process allowed for accumulating experience while executing the 225 predefined hints for each query. In the N-chunks, we maintained the original approach of the baseline model. However, a potential issue arises when attempting to execute all 225 hints for each query in the first chunk, as it can lead to an exponential increase in training time proportional to the number of predefined hints. 

To overcome this limitation, we examined whether there were any overlapping ASPs when appending the predefined hints to the queries. On average, we found that there were around 20-30 distinct ASPs, with a maximum of approximately 40. Therefore, we applied a method to remove duplicates and executed the queries only with hints that had distinct ASPs, storing the results in the experience.

\subsection{Removing Bias in Loss Function with Q-error} \label{qerror-bias-section}

Loss functions commonly used in learned query optimization cost models are the Mean Squared Error(MSE) loss. However as it takes the magnitude of all queries into account, naturally the queries with bigger running time heavily influence the model. With the purpose of managing regressions for long and substantial queries, MSE loss would be the primary go-to loss function. 

However, as an observation of our previous experiments (Figure \ref{qerror-bias-figure}), there was a performance gap between queries with short cpu time and those with long cpu time. In our first experiment, 21.4\% of former (i.e. short) queries improved while 26.3\% of latter (i.e. long) queries improved. Naturally, the model already finds substantial better plans for short-cpu-time queries, but matching the ratio of the two can be an important improvement point. If the model works to make better predictions only for long-cpu-time queries as the previous model with MSE does, then the value of using the model for other short-cpu-time queries will diminish. The more similar the ratio of the two, the fairer prediction becomes possible no matter what queries actually come in, which means the model becomes more practical and valuable. \\

Therefore, we adapted a revision of a new loss function, \textit{Q-error}. Q-error, a popular evaluation metric in the context of cardinality estimation\cite{b22, b23, b24}, mediates queries of all costs (i.e. rewards), disregarding the magnitude of queries as it is somewhat of a ratio. Q-error is effective in that it is able to cope with both overestimation and underestimation of query execution times. Therefore, we take the q-error for CPU time and subtract 1 from it, to make a perfect prediction close to 0(Equation 1). However, for convenience, it is replaced by the name Q-error. 
\sloppy
\\
\begin{multline}
\scalebox{0.8}{%
$ Q-error = \sum_{1}^{n}(max(\frac{Predicted\ CPU\ time_i}{Real\ CPU\ time_i}, \frac{Real\ CPU\ time_i}{Predicted\ CPU\ time_i})-1) $} \\
\scalebox{0.8}{$ , 0\le i\le \# experience $}
\end{multline}

\section{Learning Requirements}
In this section, we discuss the main environments and preparations necessary to set up and configure our model aligned with the expert optimizer. The characteristics of the datasets that we used for shaping the value network are described in Section \ref{Dataset} and then in Section \ref{environment-settings}, the system environments are briefly introduced. We next address the challenges that arose in integrating a learned model into the expert optimizer, specifically in reducing latency to make prediction performance comparable with the HANA engine in Section \ref{soft-landing-section}. 

\subsection{Dataset} \label{Dataset}

\begin{table}[]
\begin{tabular}{@{}l|l|l|l@{}} \midrule
\textbf{Workload} & \multicolumn{1}{c|}{\textbf{\# Columns}} & \multicolumn{1}{c|}{\textbf{\# Tables}} & \multicolumn{1}{c}{\textbf{\begin{tabular}[c]{@{}c@{}}\# Records of table \\ with the most records\end{tabular}}} \\ \hline
JOB               & 60                                       & 21                                      & 36M                                                                                                               \\
JOB-Light         & 8                                        & 6                                       & 36M                                                                                                               \\
JOB-LR  & 13                                       & 6                                       & 36M                                                                                                               \\
JOB-E      & 41                                       & 16                                      & 36M                                                                                                               \\
JOB-M             & 16                                       & 16                                      & 36M                                                                                                               \\
TPCH SF1          & 53                                       & 8                                       & 6M                                                                                                                \\
TPCH SF10         & 53                                       & 8                                       & 60M                                                                                                               \\
TPCH SF100        & 53                                       & 8                                       & 600M                                                                                                              \\
JCCH              & 53                                       & 8                                       & 60M                                                                                                               \\
TPCDS SF1         & 248                                      & 24                                      & 3M                                                                                                                \\
TPCDS SF10        & 248                                      & 24                                      & 30M                                                                                                               \\
STACK             & 39                                       & 10                                      & 92M            \\ \bottomrule                                                                                                
\end{tabular}
\newline
\caption{\label{workloads-table}Full list of workloads mainly used to comprise the training set. Number of records of table with the most records nearly represents scale factor, which is taken into account when splitting workloads for Ensemble. JOB-LR and JOB-E each stand for JOB-Light-Ranges and JOB-Extended. }
\end{table}

Our training dataset were extracted from the workloads listed in Table \ref{workloads-table}. 

JOB are original 113 queries from IMDB dataset, an augmentation of the Join Order Benchmark\cite{b15}. Similarly, JOB-light is a benchmark popularly used in cardinality estimation studies. It contains 70 queries with 6 table-schema and each query joins 2-5 tables. JOB-light-ranges contain more complex filters are added based on JOB-light while JOB-M\cite{b16} contains more complicated join conditions and each query joins 2–11 tables. JOB-Extended \cite{b17} use same relations and different sematics compared to JOB. We modified some queries so that HANA database can compile them successfully. TPCH\cite{b18} and TPCDS\cite{b19} are a decision support benchmark with each of 22 and 99 queries. There are various scaling factors for TPCH and TPCDS benchmark. For example, SF10 consists of the 10 times larger base row size than SF1, since different scaling factors represent different dataset sizes. JCCH\cite{b20} has a skewed version of TPCH SF10 schemas but uses the same TPCH queries. STACK is real-world dataset and workload from Bao \cite{b3}. 

\subsection{Environment settings} \label{environment-settings}

\textbf{System specification} All experiments were performed on a corporate server, using a machine of 2 NVIDIA TITAN Xp GPUs coupled with 22 CPU cores and 1009 GB of DDR4 RAM. 

\textbf{System configurations} To ensure the consistency of query execution environment and performance, 25 system configurations were altered and fixed(e.g. such as enforcing execution processing engine version, compile time sampling size and to collect and use runtime feedback). 

\textbf{Why CPU instead of elapsed time?} When a query is executed, it is run in parallel over several threads. The number of threads used to execute the query varies for every query or situation, hence it fails to be a coherent value to predict. Therefore, we prefer to work with CPU time which should be the total time taken to execute each query, taking every thread into consideration.

\textbf{Stable execution time} \label{stable-execution-time} CPU average execution time can often be fluctuant depending on how burdened the machine is in the status quo, how many jobs are running and what version of DB build it is using. Also, the initial executions of queries are much slower than the latter comings as the plan cache is non-existent for unseen queries. Therefore to achieve a stable, consistent CPU time we clear the cache before the initial execution and warm up the cache 3 times. Then we take the following 3 executions and take the average to be the reward of each query which leads CPU execution times to be coherent and viable to construct our value model with. 

\subsection{Soft landing and Parallel compilation} \label{soft-landing-section}
\subsubsection{\textbf{Parallel compilation}} To predict the optimal plan out of 225 hintsets, we first need to compile and capture all ASPs. To reduce the redundancy in compilation time and prevent regression, the abstract sql plans for each hintset are run in parallel. As we work with large amounts of threads, we ensure that there are no abnormally hanging threads or database connections by limiting the max pool size to 225. 

\subsubsection{\textbf{Soft landing with timeout}} One of the notable challenges we faced during building the model was that query execution can be extremely slow. Sometimes, it is the matter of inefficient plan generation, while a lot of the times, it is due to the fact that the query itself is too heavy. Out of the queries we executed with the default plan generated from the expert optimizer, the maximum elapsed execution time was approximately 130 minutes with the average elapsed execution time being 4 seconds. Considering that we execute each query 6 times to retrieve the warmed-up average of the execution time, only the data extraction itself would overpower the training time of the model itself. 

This is problematic in that to collect data, we execute queries in real time with the database engine, sufficient enough to train a deep reinforcement learning model. Another point to keep in mind is that we also need to extract catastrophic plans and its reward time to learn its patterns and prevent the model to choose such plans. 

Therefore as a soft landing mechanism, we opt for timeouts to limit the execution of queries with disastrous plans. Queries that take longer than the preset connection timeout will be terminated and not be included in the experience. Most common values used for timeouts are 10 and 20 seconds. 

\textbf{Does this hinder the learning of queries that take longer execution? }

We observed that the prediction does not degrade even for queries and plans that take longer than the connection timeout. 10 seconds of time limit provides a window just enough to capture the relationship of complex plans but not elongate the query execution phase excessively. Once the pattern of the plan tree is learned, size and cost are taken into context together to predict the CPU execution time. \\ 

\section{Related Works}
The modern DBMS optimizer uses the cost model to select the best plan among the best query execution plans (QEPs). A key part of query optimization is to improve cardinality estimation, and the latest research is trying to apply machine learning to these cost models. Recently, many studies have been conducted on \textbf{1) how to improve cardinality estimation using supervised and unsupervised learning} or \textbf{2) how to generate query plans after learning using cost model or runtime.} 

Many studies have been conducted to increase the acuity of cardinality estimation as a way to improve runtime. Rong et al. stated that FLAT \cite{b11} has emerged to solve the existing problem of cardinality estimation, slow probability computing. It has fast probability computing, light weight for model size, and pretty accurate estimation quality. Based on IMDB benchmark, query execution time has improved by 12.9\%. 

The query-driven method has limitations in that it has dependency on training query and the quality of training query affects the estimation quality. The data-driven method has fewer limitations and higher adaptability compared to query-driven method. For this reason, FACE \cite{b12} used data-driven methods and proposed the model by capturing the dependency of column and large domain size for high accuracy, high inference efficiency, and achieving lightweight model size. It improved latency and memory usage compared to previous research, on the contrary, there are still constraints for acyclic joins or other predicates. 

Reference \cite{b13} implemented an unified deep autoregressive model that attempts to narrow the gap between data-driven and query-driven model. It operates learning joint data distribution using both data and query workload. There is improvement in that it supports group-by-clauses and learns both data and query, but constraints still exist for various joins.

LEO \cite{b14} is one of the earliest applications, and it is operated by estimating cardinality with a query execution plan. It develops by comparing the predicted value to the actual value of each step of the actual QEP and giving feedback to learn from the past mistake. Stillger et al. asserted that this model showed successful performance for similar queries. 

NEO \cite{b4} is the first fully learned system for creating QEPs. It predicts the smallest cost through featurization with the value among the various QEPs, throwing a selected plan, and using a latency with the actual execution for experience. However, the model has limitations in that it has database dependency.  

Marcus et al. \cite{b3} implemented a model which can create QEPs by attaching predefined hints respectively and predicts hints that have the smallest cost among the plans using the TCNN structure. It has the advantage of being able to operate by attaching this model to the existing optimizer, and there is no database dependency. NEO gives a plan as output, however, Bao returns a hint that can have good performance for a specific query as an output. Unlike NEO, the input requires cardinality and cost, and the information must be obtained from the existing optimizer such as HANA. In addition, Bao is a workload-driven method, so the query execution of a large set of training queries is required.

\section{Experiments and Results}
Our key evaluation is whether our enhanced model successfully removes bias of the baseline model to prefer patterns of no-hint plans generated from underlying expert optimizer (Section \ref{Entire Hintsets}) and performs better on mixed-workload environment with ensemble method (Section \ref{hypothesis}). Here we present and visualize the key findings of experiments mentioned above.

\subsection{Results for Removing Bias}

\begin{figure}[htbp]
\centerline{\includegraphics[width=8cm]{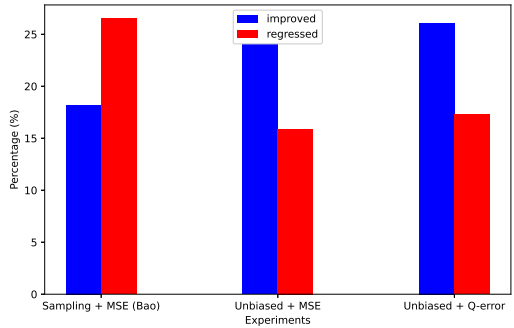}}
\caption{Improved and Regressed percentage of each experiment: Sampling + MSE denotes original baseline model.}
\label{improved-regressed-figure}
\end{figure}

We aimed to increase the possibility of finding better plans than those preferred by the expert optimizer by addressing the bias issue. 
\subsubsection{\textbf{Utilizing entire hintsets}} To achieve this, we first adopted the approach of utilizing all hint sets, as introduced in Section \ref{Entire Hintsets}. The experimental results of the original approach - which stacks initial data with no-hint - are well represented in the first row of Table \ref{entire-bias-results}. The "No-hints" category indicates that using the existing optimizer without any additional hints was the best approach, with a predicted ratio of 42.4\% while both the "Improved" and "Regressed" categories had a ratio of 18.2\% and 26.5\%. 

\begin{table}[]
\centering
\begin{tabular}{|l|l|l|l|}
\hline
Experiment & \multicolumn{1}{c|}{\# Improved} & \multicolumn{1}{c|}{\# Regressed} & \multicolumn{1}{c|}{\# No-hints} \\ \hline
Baseline Model   & 18.2\%                        & 26.5\%                         & 42.4\%                        \\ \hline
BitE(Unbiased)   & 24.3\%                        & 15.9\%                         & 42.5\%                        \\ \hline
\end{tabular}
\newline
\caption{No-hint predictions and queries with less of an 10\% CPU time change are excluded from improved and regressed counts. 18.2\% of the queries have shown more than 10\% better performance with the hint our learned model has selected compared to HANA default plan. }
\label{entire-bias-results}
\end{table}

With entire hint sets, the ratio of "no-hint" increased from the original 42.4\% to 42.5\%, a reduction of approximately 0.1\%. This might question the effectiveness of reducing bias based on this outcome. However, regression decreased from 26.5\% to 15.9\%, a reduction of approximately 10.6\%. This outcome can be explained by comparing the actual execution times which shows it is evident that there are queries where 'no hint' emerges as the optimal choice. Also, the percentage of queries that were improved increased from 18.2\% to 24.3\%, a gain of approximately 6.1\%. By expanding the search space for queries that were initially selected with no-hint, we were able to steer the model to make more assertive decisions.

Furthermore, it can be observed that queries that previously made incorrect selections undergo a transition towards making appropriate selections with bold choices. As shown in Figure \ref{improved-regressed-figure}, enhancements were seen in both the number of improved and regressed queries. This suggests that by reducing the tendency of the model to learn patterns favored by the expert optimizer, we were able to discover good plans that were previously overlooked due to the bias existing in the expert optimizer and conduct more aggressive exploration, which involves exploring a broader search space instead of relying on plans selected by the expert optimizer. 

\begin{figure*}[htbp]
\begin{subfigure}[b]{0.3\textwidth}
\includegraphics[width=\textwidth]{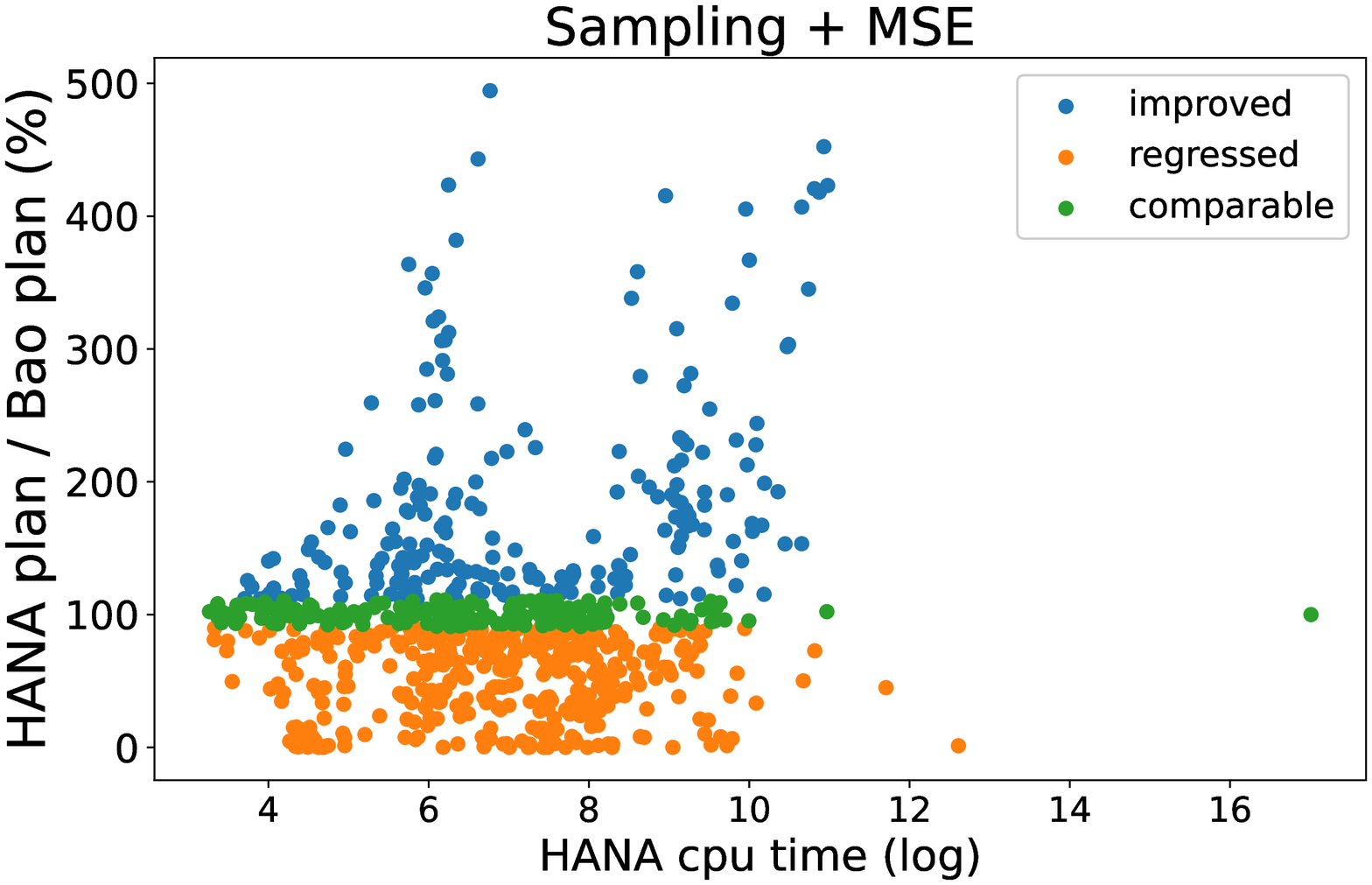} \caption{Baseline model(Sampling + MSE)} \label{fig:3a}
\end{subfigure}%
 \hfill
\begin{subfigure}[b]{0.3\textwidth}\includegraphics[width=\textwidth]{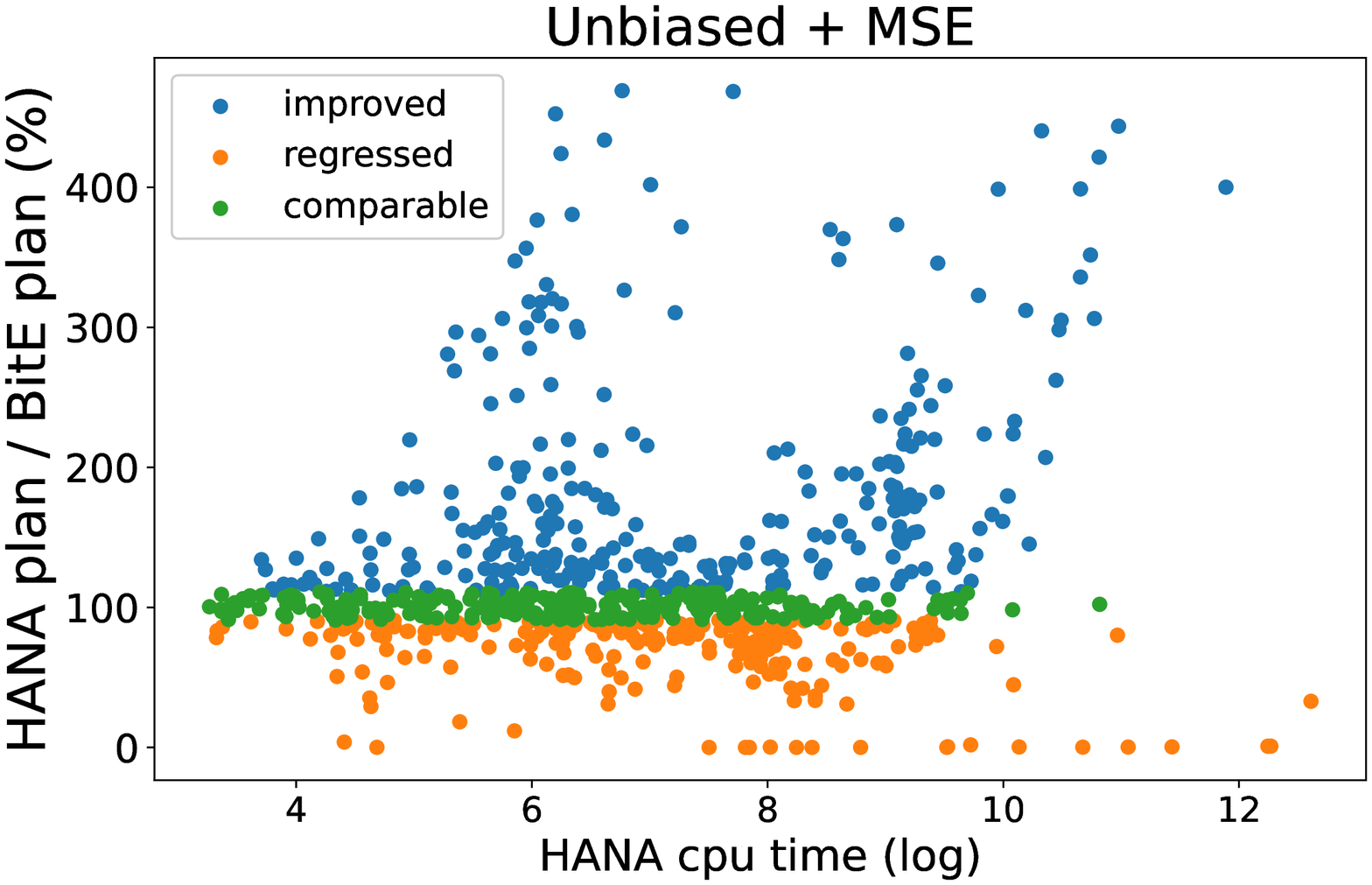} \caption{Unbiased + MSE} \label{fig:3b}
\end{subfigure}%
 \hfill
\begin{subfigure}[b]{0.3\textwidth}\includegraphics[width=\textwidth]{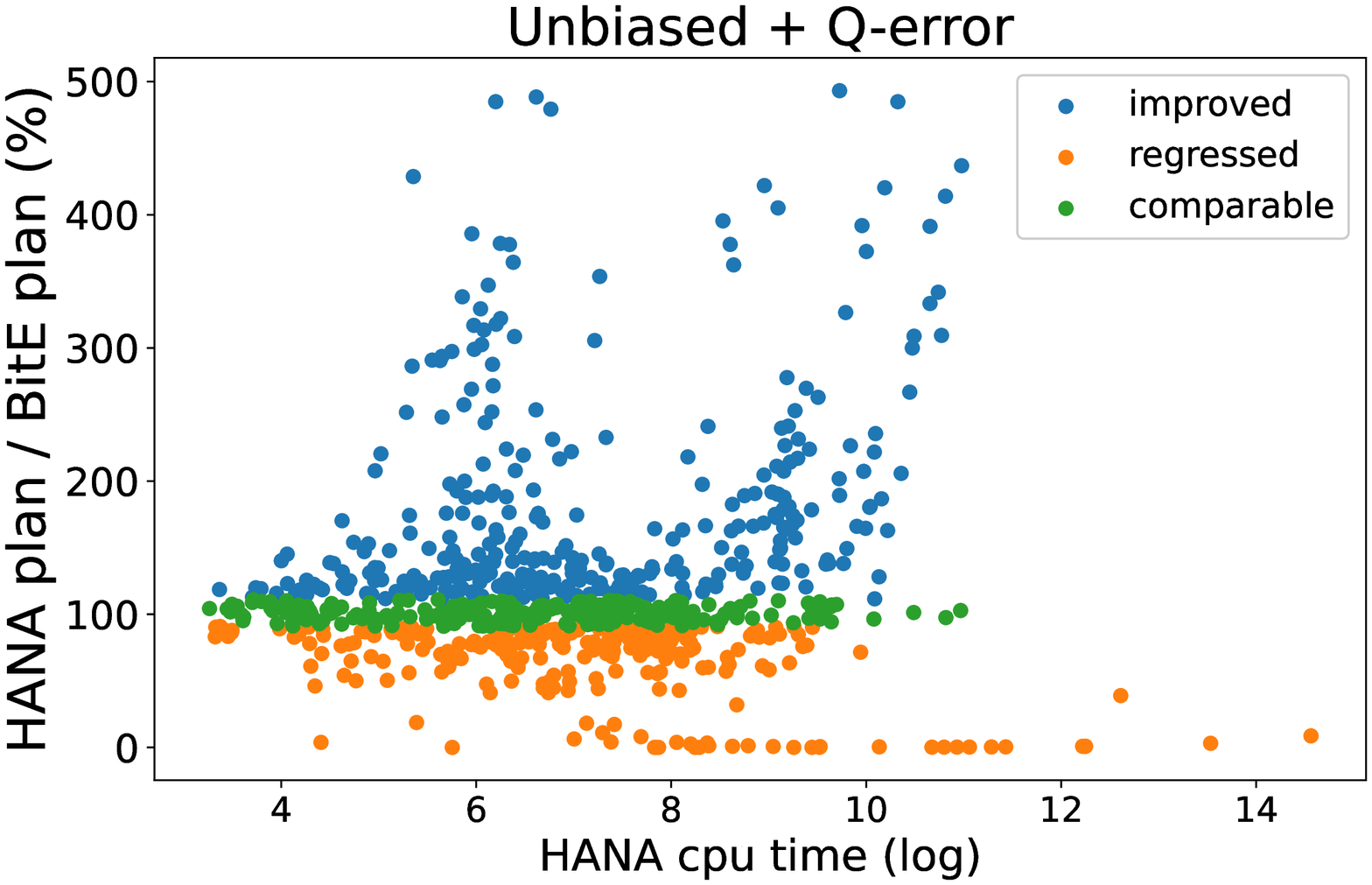} \caption{Unbiased + Q-error } \label{fig:3c}
\end{subfigure}%
\caption{ Cpu time comparison between HANA plan and the plan selected by the learned model in the experiments 
}
\label{cpu-comparison-figure}
\end{figure*}

\subsubsection{\textbf{Q-error loss function.}} 
\begin{figure}[htbp]
\centerline{\includegraphics[width=9cm]{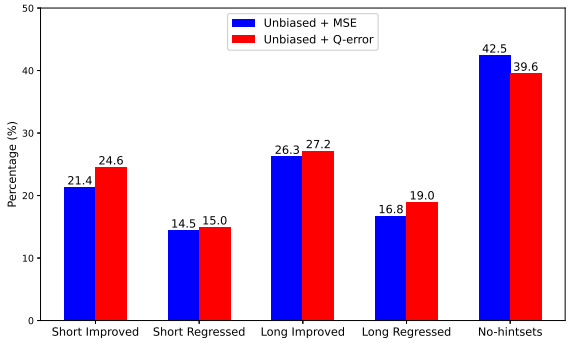}}
\caption{Improved and Regressed percentage of each experiment. Short and Long each denote the set of queries divided by the average CPU execution time.}
\label{qerror-bias-figure}
\end{figure}

Nevertheless, a problem exists that the prediction results were better for queries with long cpu time. To solve this, Q-error was introduced to our losss function.
A remarkable point of the results of using Q-error is that improvement of short-cpu-time queries have increased to a similar degree of long-cpu-time-queries. As shown in Figure \ref{qerror-bias-figure}, the difference between Short improved and Long improved was 4.9\% before, and with Q-error the difference of the two decreased to 2.6\%. This indicates that Q-error fairly improves prediction on all queries and the distribution becomes more even. Figure \ref{fig:3b} and \ref{fig:3c} visualize the results again with more improved queries of short-cpu-time queries. Also the results were meaningful even when viewed as a whole(Figure \ref{improved-regressed-figure}) rather than looking at short-time-queries and long-time-queries separately. Improvement of all queries increased compared to MSE loss function.

\begin{figure}[ht]
\begin{subfigure}[b]{4.2cm}
\includegraphics[width=4.2cm]{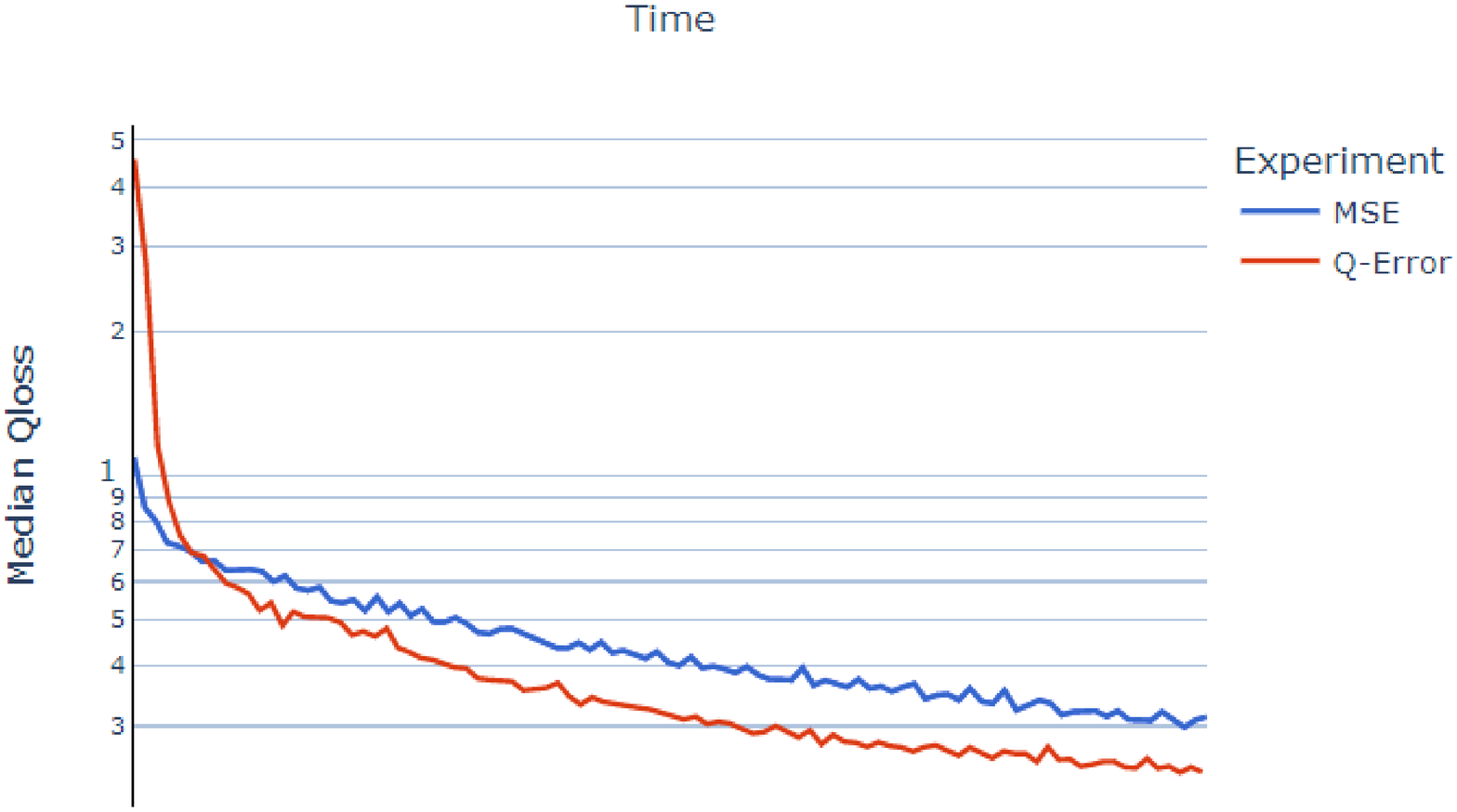} \caption{Median Q-Error} \label{fig:2a}
\end{subfigure}%
 \hfill
\begin{subfigure}[b]{4.2cm}\includegraphics[width=4.2cm]{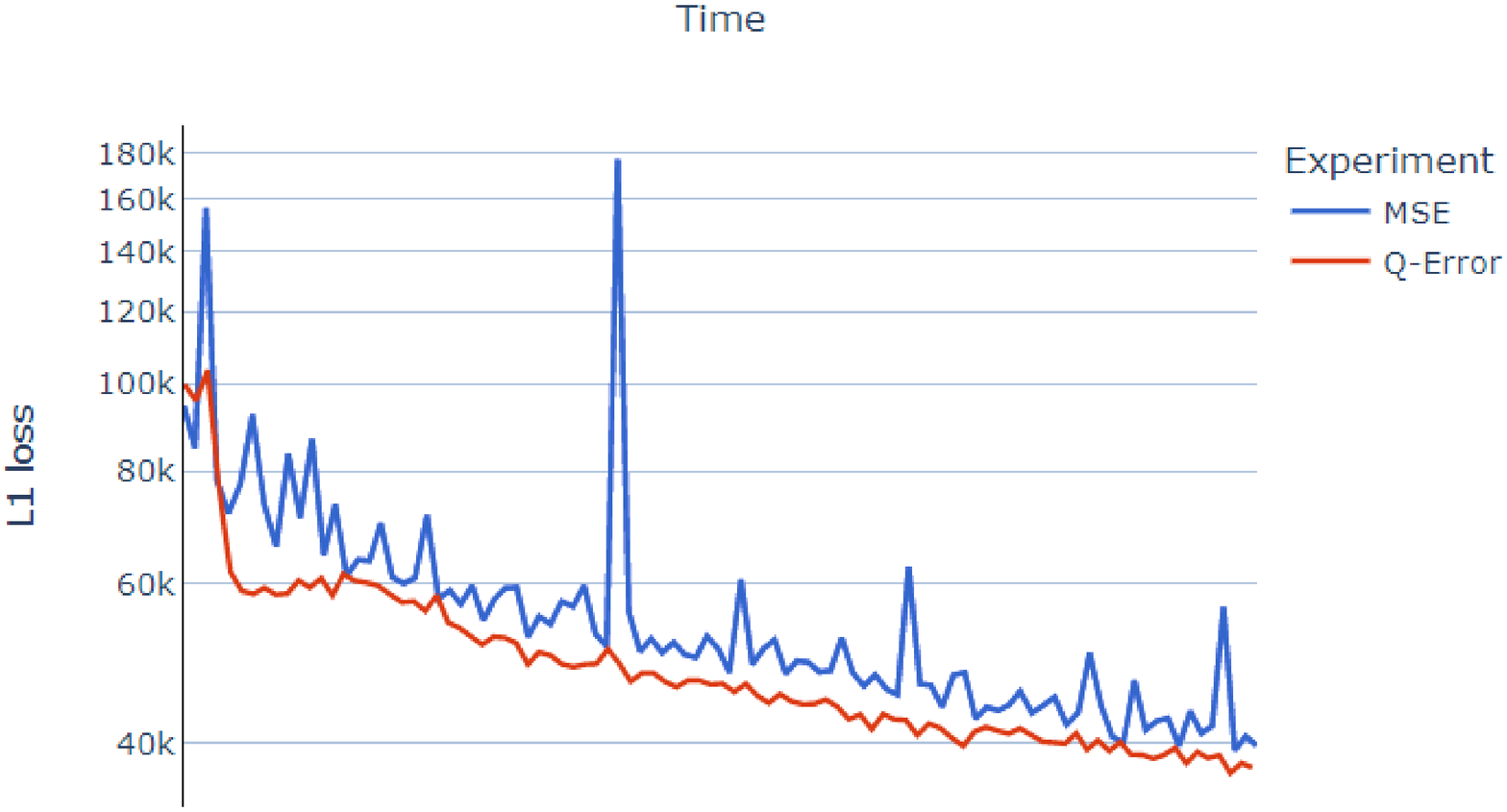} \caption{Mean L1 Loss} \label{fig:2b}
\end{subfigure}%
\caption{ Median Qloss and L1 loss comparison between model with MSE and Q-error loss function. Short-cpu-time (e.g. Short) queries and long-cpu-time (e.g. Long) queries are divided based on median value of HANA cpu time for pre-extracted queries.}
\label{qloss-figure}
\end{figure}

Figure \ref{qloss-figure} show the Qloss of two experiments, our previous experiment with MSE loss function and our new experiment with Q-error loss function measured in training phase. As we trained the model with mean Qloss to reduce tail latency, mean Qloss decreased by about 10\% compared to the model with MSE loss. Since the mean value reflects and takes into account predictions of all queries, Qloss of quantile 90 also decreased by 30\%, which indicates Q-error improved the tail performance. Extra median Qloss and L1 loss were measured. Q-error performed better than MSE for median Qloss by about 10\% and L1 loss also reduced by about 30\%.  

Another point is that training time it takes for median Qloss to converge to the same percentage – say 0.298, the converged Qloss in MSE version - have shortened by about two times in Q-error version. Additionally, with Q-error, it takes about two and a half hours of training time, and during that time, 20,000 records are processed for median Qloss to reach 0.14 and mean Qloss to reach a level of 0.6. 

\subsection{Ensemble Learning results} \label{ensemble-results-section}

Here, we discuss and evaluate the results of the ensemble run in comparison with the outcomes in Section \ref{hypothesis}. For the ensemble runs, we adopt Q-error, proven to show better efficiency, to tune our model and learn from the unique abstract sql plans among 225 hintsets, while we take on utilizing separate models for Light and Heavy workloads. As proposed in Table \ref{split-table}, the training of the Light and Heavy models are performed on the predetermined workloads based on their database metadata and complexity. The same logic applies for the predictions evaluation, where a query is put in and at the forefront, its complexity is classified by its workload and is sent into the ensemble model. 

\begin{figure}[htbp]
\centerline{\includegraphics[width=0.5\textwidth]{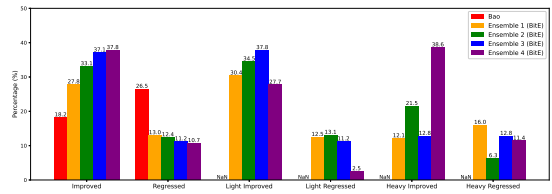}}
\caption{Improved and regressed percentage of 4 experiments.}
\label{ensemble-results-figure}
\end{figure}

\textbf{Is using table statistics and metadata the optimal way?} Compared to the results of Ensemble 1 and 2 which were based more on our experience of latency during real execution, Ensemble 3 and 4 were divided by a clear and objective rule using the three factors(Section \ref{intro}). Its results both excel in the improved and regressed queries in comparison with the former two experiments. Figure \ref{ensemble-results-figure} also illustrates the performance of each of the separate complexity models which show that Ensemble 3 shows a powerful improvement for Light workloads meanwhile Ensemble 4 has a surprising low percentage of Light regression queries with a strong improvement in the Heavy improved queries. 

\begin{figure*}[htbp]
\begin{subfigure}[b]{0.3\textwidth}
\includegraphics[width=\textwidth]{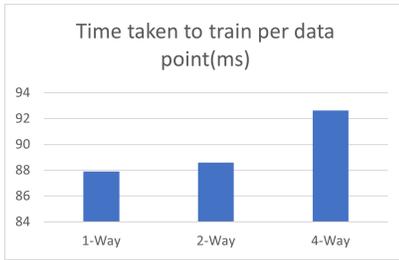} \caption{Time taken to train learned model per data point} \label{fig:6a}
\end{subfigure}%
 \hfill
\begin{subfigure}[b]{0.3\textwidth}\includegraphics[width=\textwidth]{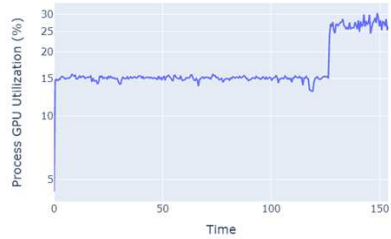} \caption{Process GPU utilization of Ensemble model(i.e. two-way model)} \label{fig:6b}
\end{subfigure}%
 \hfill
\begin{subfigure}[b]{0.3\textwidth}\includegraphics[width=\textwidth]{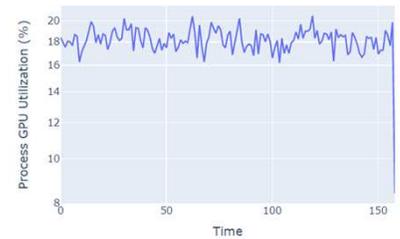} \caption{Process GPU utilization of a one-way model(i.e. baseline model)} \label{fig:6c}
\end{subfigure}%
\caption{Comparison of resource usage for different models. (\ref{fig:6a}) Training time for n-way models. N represents how many classifications of workloads exist to consist the Ensemble model. (\ref{fig:6b}) Process GPU Utilization for training Light and Heavy Ensemble model. The latter increase denotes utilization for Heavy (\ref{fig:6c}) Process GPU utilization of the baseline model}
\label{n-way-figure}
\end{figure*}

\begin{figure}[htbp]
\centerline{\includegraphics[width=9cm]{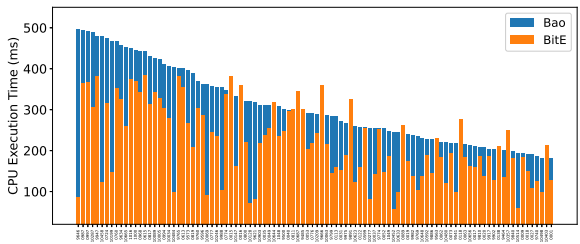}}
\caption{Comparison of CPU execution time with plans produced by Bao and BitE(Ensemble 3). The difference between the query latency is shown for a subset of JOB\_light\_ranges queries.}
\label{cpu-diff-figure}
\end{figure}

\textbf{Do BitE have an edge over the original baseline model?} Figure \ref{ensemble-results-figure} shows that no matter how complexity for the ensemble methods is divided, its results are significantly better both in the number of improved and regressed queries out of 20,000 records that cover all mentioned workloads. It also illustrates that even when the balance between the number of workloads that consist of the Light and Heavy components is unequal, such as Ensemble 3 and Ensemble 4 (according to Table \ref{split-table}), the results are far acceptable than the original model which is a melting pot of all workloads. This proves that the segmentation of the entire model by grouping workloads that have similar complexity is significantly efficient.

\textbf{Memory and GPU usage} Apart from the inference results, an important question of resource usage is left to be addressed as well. If the training is done separately for the Light and Heavy complexity, does it require more resources than simply constructing one model that covers all workloads? Does dividing into more ensemble models always guarantee better performance? We answer this question by looking into the time spent and GPU utilization during the training. Figures 6.a shows the time taken to train the model per data point each for the models that are divided into 1, 2 and 4 types of workload ensemble models. The more ensemble models we distinguish, the longer it takes to converge(Figure \ref{fig:6a}).

This leads to an increase in GPU utilization. Figure \ref{fig:6b} and \ref{fig:6c} show that the average process GPU utilization for Ensemble training of 17.1\% is slightly lower than for the single model of 17.5\%. At the same time, the Ensemble training displays difference in training time of 20,000 records with a reduction of 3\% of the total training time in comparison to the single model. Hence we far-outperform the original single model with our proposed ensemble method with less or at least, a comparable level of resource usage.

\textbf{Safe Tail-latency Plans} Figure \ref{cpu-diff-figure} depicts a detailed understanding of how good BitE really is-lower execution time being better. It not only manages to reduce the tail-latency in execution time by avoiding catastrophic choices but it also easily surpasses the performance of the baseline model in most queries of the JOB light ranges workload. Out of 100 target queries, 87\% queries successfully find better plans to enhance and accelerate query execution with BitE. For the rest of the 13\%, the queries do fall into regression compared with Bao. However, the found hintsets are still acceptable as execution time increases by, at most, only 25\% while some of the improved plans tune and reduce their execution time almost to a striking 80\%.

In summary, BitE succeeds to find better plans and outperform the existing learned query optimization methods by taking advantage of the workload characteristics and tuning certian elements to remove underlying bias. The Ensemble methods are simple, reproducible yet not expensive to implement on top of any expert database optimizer.

\section{Conclusion and Future Work}
In this paper, we presented novel methods to accelerate query optimization and enhance its performance. In a mixed-workload environment, the ensemble model is divided and trained based on the complexity and characteristics of different workloads. Along the way, we introduced three novel propositions and ideas of expanding hintsets, deviating from the default plan bias and revising the loss function to generalize query optimization for queries with varying weight. This provides a more acute grasp of query optimization and efficiently outperforms the base work provided by the baseline model. 

In the future, we find it an interesting direction to expand the search space of optimal plans with additional hintsets and to narrow down the traits of queries that some hintsets are more likely to apply to. Furthermore, the number of unique plans (as certain hints may not apply to queries) per query or workload can be considered as another factor to determine the complexity and build the ensemble model.

%%
%% The next two lines define the bibliography style to be used, and
%% the bibliography file.
\bibliographystyle{ACM-Reference-Format}

\end{document}